\documentclass[12pt]{article}
\usepackage{makeidx}
\usepackage{amsmath}
\usepackage{amsfonts}
\usepackage{amssymb}
\usepackage{epsfig}
\usepackage[cp1251]{inputenc}

\newcounter{defin}
\newcounter{lemma}
\newcounter{theorem}
\newcounter{proposition}
\newcounter{example}
\newenvironment{lemma}{\par\refstepcounter{lemma}     \textbf{Lemma \thelemma.} }{\rm\par}
\newenvironment{theorem}{\par\refstepcounter{theorem}     \textbf{Theorem \thetheorem.}\ }{\rm\par}
\newenvironment{proposition}{\par\refstepcounter{proposition}     \textbf{Proposition \theproposition.}\ }{\rm\par}

\begin{document}

\title{The structure of general quantum Gaussian observable}
\author{A.S. Holevo \\
Steklov Mathematical Institute, RAS, Moscow, Russia}
\date{}
\maketitle

\begin{abstract}
The structure theorem is established which shows that an arbitrary
multi-mode bosonic Gaussian observable can be represented as a combination
of four basic cases, the physical prototypes of which are homodyne and
heterodyne, noiseless or noisy, measurements in quantum optics. The proof establishes connection between the
description of Gaussian observable in terms of the characteristic function and
in terms of density of the probability operator-valued measure (POVM) and
has remarkable parallels with treatment of bosonic Gaussian channels in
terms of their Choi-Jamiolkowski form. Along the way we give the
``most economical'', in the sense of minimal dimensions of the quantum ancilla, construction of the Naimark extension of a general
Gaussian observable.
It is also shown that the Gaussian POVM has bounded operator-valued density
with respect to the Lebesgue measure if and only if its noise covariance matrix is
nondegenerate.
\end{abstract}

\section{Introduction}

The most general definition of Gaussian observable for multi-mode bosonic
continuous-variable systems was formulated in the book \cite{QSCI}, basing
on important special cases previously considered by different authors (see
e.g. the book \cite{aspekty} and references therein). There are basic
physical prototypes -- one is approximate or exact position measurement, the
other is approximate joint position-momentum measurement; in quantum optics
these correspond to (noiseless or noisy) homodyne vs (vacuum or thermal noise)
heterodyne measurements of the radiation field quadratures \cite{caves}. In
this paper we establish the structure theorem which shows that an arbitrary
multi-mode bosonic Gaussian observable can be represented as a combination
of these four basic types. The proof establishes connection between the
description of Gaussian observable in terms of characteristic function and
in terms of density of the probability operator-valued measure (POVM) and
has remarkable parallels with treatment of bosonic Gaussian channels in
terms of their Choi-Jamiolkowski form \cite{cj}. Along the way we give the
``most economical'' construction of the Naimark extension of a general
Gaussian observable, in the sense of the minimal dimensions of the quantum ancilla.
It is also shown that the Gaussian POVM has bounded operator-valued density
with respect to the Lebesgue measure if and only if its noise covariance matrix is
nondegenerate.

The possibility of complete description of the structure of arbitrary Gaussian observable
demonstrated in theorem \ref{t1} renews the interest to the structural analysis of the general quantum Gaussian
channels. That problem is much more involved (cf. \cite{wolf}) and was successfully solved only for
the gauge-covariant channels entailing resolution of the long-standing ``Gaussian maximizer'' problem for the
classical capacity of such channels \cite{ghg}.  In our classification of Gaussian observables
we do not impose the gauge-covariance, but mention in passing that the gauge-covariant Gaussian observables
fall into our type 1. The classical capacity of the general type 1 Gaussian observables was computed in \cite{acc-noJ}
under certain ``threshold condition''. Notably, the ``Gaussian maximizers'' and hence the (unassisted)
classical capacity are still open problems for general type 2 Gaussian observables which are in a sense opposite to the
gauge-covariant ones. 

\section{Gaussian Observables}

Assume that we have two systems $A$ and $B$, the system $A$ is quantum
bosonic with $s$ degrees of freedom (modes) and the system $B$ is classical
and described by an $m$-dimensional linear space $Z_{B}=\mathbb{R}^{m}.$ Let
$(Z_{A},$ $\Delta _{A})$ be the symplectic vector space underlying the system $A,$
which consists of vectors\footnote{%
We denote by $^{t}$ transposition of vectors and matrices.} $\,z_{A}=[x_{1},y_{1},\dots
,\,x_{s}\,,\,y_{s}]^{t} $ , and equipped
with the symplectic form%
\begin{equation*}
\Delta _{A}\left( z,z^{\prime }\right) =z^{t}\Delta z^{\prime };\quad \Delta
=\mathrm{diag}\left[
\begin{array}{cc}
0 & -1 \\
1 & 0%
\end{array}%
\right] _{j=1,\dots ,s}.
\end{equation*}%
We denote by $W_{A}(z_{A})=\exp iR_{A}z_{A}$ an irreducible Weyl system in a
Hilbert space $\mathcal{H}_{A}$ , where $R_{A}=[q_{1},p_{1},\dots
,\,q_{s}\,,\,p_{s}]$ are the canonical observables of the system $A.$ The
Weyl canonical commutation relations imply%
\begin{equation}
W_{A}(z_{A})W_{A}(z_{A}^{\prime })W_{A}(z_{A})^{\ast }=\exp \left( -i\Delta
_{A}\left( z,z^{\prime }\right) \right) W_{A}(z_{A}^{\prime }).  \label{weyl}
\end{equation}

Let $M$ be an observable in $\mathcal{H}_{A}$ with the outcome set $Z_{B},$
given by the probability operator-valued measure (POVM) $M(d^{m}z)$. The
observable is completely determined by the \emph{operator characteristic
function }(see \cite{QSCI}):
\begin{equation*}
\phi _{M}(w)=\int_{Z_{B}}\mathrm{e}^{i\,z^{t}w}M(d^{m}z),\quad z,w\in Z_{B}.
\end{equation*}%
It has the following characteristic properties : 1) $\phi _{M}(0)=I_{A};$ 2)
$w\rightarrow \phi _{M}(w)$ is continuous in the weak operator topology; 3)
for any choice of a finite subset $\left\{ w_{j}\right\} \subset Z_{B}$ the
block matrix with operator entries $\phi _{M}(w_{j}-w_{k})$ is nonnegative
definite.

Observable $M$ will be called \emph{Gaussian} if its operator characteristic
function has the form
\begin{eqnarray}
\phi _{M}(w) &=&W_{A}(Kw)\exp \left( il^{t}w-\frac{1}{2}w^{t}\alpha w\right)
\label{ocf} \\
&=&\exp \left( i\left( l^{t}+R_{A}K\right) w-\frac{1}{2}w^{t}\alpha w\right)
,  \notag
\end{eqnarray}%
where $l\in Z_{B},$ $K:Z_{B}\rightarrow Z_{A}$ is a linear operator (real $%
2s\times m-$matrix) and $\alpha $ is a real symmetric $m\times m-$matrix.
The triple $(l,K,\alpha )$ defines parameters of the Gaussian observable.
The parameter $l$ can be made zero by corresponding shift of observable
values $z,$ and in what follows without loss of generality we assume $l=0.$
Then (\ref{ocf}) becomes%
\begin{equation}
\phi _{M}(w)=\exp \left( iR_{A}Kw-\frac{1}{2}w^{t}\alpha w\right) .
\label{ocf0}
\end{equation}%
A necessary and sufficient condition for relation (\ref{ocf}) to define an
observable is the matrix inequality \cite{QSCI}
\begin{equation}
\alpha \geq \pm \frac{i}{2}K^{t}\Delta K.  \label{mudelta}
\end{equation}

In particular, sufficiency of the condition (\ref{mudelta}) can be
established by using a construction of the Naimark extension of observable $%
M $, which we give here in the \textquotedblleft most
economical\textquotedblright\ version, in the sense of the minimal number of
modes of the quantum ancilla.

We denote $\Delta _{K}=K^{t}\Delta K,$ which is a skew-symmetric $m\times m-$%
matrix of commutators between the components of the vector operator $%
R_{K}=R_{A}K.$ \ We denote by $r_{\Delta _{K}}$ the rank of $\Delta _{K}$
which is necessarily even, $r_{\Delta _{K}}=2s_{1},$ and by $r_{\alpha }$
the rank of the matrix $\alpha $.

\begin{theorem}\label{t1}
\label{gausnaimark} \emph{Assume the condition (\ref{mudelta}), then there
exists an ancillary Bosonic system (ancilla) $C$ with $%
s_{C}=r_{\alpha }-r_{\Delta _{K}}/2$ quantum modes in the space $\mathcal{H}_{C}$,
built on a symplectic space $\left( Z_{C},\Delta _{C}\right) $ , a Gaussian state $%
\rho _{C}$ \ in $\mathcal{H}_{C}$, and a projection-valued measure $%
E_{AC}(d^{m}z)$ in the space $\mathcal{H}_{A}\otimes \mathcal{H}_{C}$ \ such
that
\begin{equation}
M(U)=\mathrm{Tr}_{C}\left( I_{A}\otimes \rho _{C}\right) E_{AC}(U),\quad
U\subseteq Z_{B},  \label{nai}
\end{equation}%
where $I_{A}$ is unit operator in $\mathcal{H}_{A}.$ Namely, $\rho _{C}$ \
is centered Gaussian state with the covariance matrix $\alpha _{C}$
satisfying
\begin{equation}
\quad K^{t}P^{t}\Lambda \alpha _{C}\Lambda PK=\alpha ,  \label{proj}
\end{equation}%
where $\Lambda $ is involution in $Z_{C}$ such that $\Lambda \Delta
_{C}\Lambda =-\Delta _{C},$ and $P$ is a projection; the projection-valued
measure $E_{AC}$ is the joint spectral measure of the commuting selfadjoint
components of the vector operator}
\begin{equation}
X_{B}=R_{A}K\otimes I_{C}+I_{A}\otimes R_{C}\Lambda PK.  \label{selfa}
\end{equation}
\end{theorem}

The main ingredient of the proof is the construction of the system $C$, of
the covariance matrix $\alpha _{C}$ of the state $\rho _{C}$ and of the
transformation $\Lambda P$ underlying the definition of the spectral measure
$E_{AC}$ , which will be given in sec. \ref{s2}. Assuming this, the
characteristic function of the observable $E_{AC}$ is
\begin{eqnarray*}
\phi _{E_{AC}}(w) &=&\int_{Z_{B}}\mathrm{e}^{i\,z^{t}w}E_{AC}\left(
d^{m}z\right) \\
&=&\exp \left( iX_{B}w\right) =\exp i\left( R_{A}K\otimes I_{C}+I_{A}\otimes
R_{C}\Lambda PK\right) w \\
&=&W_{A}(Kw)W_{C}(\Lambda PKw),
\end{eqnarray*}%
whence, denoting by $\rho _{C}$ the centered Gaussian state with the
covariance matrix $\alpha $,
\begin{eqnarray*}
\mathrm{Tr}_{C}\left( I_{A}\otimes \rho _{C}\right) \phi _{E_{AC}}(w)
&=&W_{A}(Kw)\exp \left( -\frac{1}{2}w^{t}K^{t}P^{t}\Lambda \alpha
_{C}\Lambda PKw\right) \\
&=&W_{A}(Kw)\exp \left( -\frac{1}{2}w^{t}\alpha w\right) =\phi _{M}(w),
\end{eqnarray*}%
and (\ref{nai}) follows.

Without loss of generality, we will assume that $K$ is column-independent
(in particular, $m\leq 2s$ and $K^{t}K$ is nondegenerate $m\times m-$%
matrix). This means that the components of $R_{A}K\equiv R_{K}$ are linearly
independent. Also $K$ is an injection of $Z_{B}$ into $Z_{A}$ because $Kw=0$
implies $K^{t}Kw=0$ and hence $w=0.$

General results of \cite{H-entbrch} imply that the POVM $M$ can be
represented as%
\begin{equation}
M(U)=\int_{U}m(z)d^{m}z,\quad U\subseteq Z_{B},  \label{den}
\end{equation}%
where $m(z)$ are densely defined, positive definite, in general nonclosable,
quadratic forms. When they are closable, the values of the density $m(z)$ are bounded
operators. Our analysis in section \ref{s2} will show the following
result:

\begin{proposition}
\label{p1} \emph{The condition $\det \,\alpha \neq 0$ is necessary and
sufficient for the Gaussian POVM (\ref{ocf}) to have bounded operator-valued
density.}
\end{proposition}

Meanwhile, assuming (\ref{den}) we have%
\begin{equation*}
\int_{Z_{B}}\mathrm{e}^{i\,z^{t}w}m(z)d^{m}z=\exp \left( iR_{A}Kw-\frac{1}{2}%
w^{t}\alpha w\right).
\end{equation*}%
Inverting the Fourier transform and using (\ref{weyl}), we get
\begin{eqnarray}
m(z) &=&\frac{1}{\left( 2\pi \right) ^{m}}\int_{Z_{B}}\mathrm{e}%
^{-i\,z^{t}w}\exp \left( iR_{A}Kw-\frac{1}{2}w^{t}\alpha w\right) d^{m}w
\notag \\
&=&W_{A}(K_{1}z)m(0)W_{A}(K_{1}z)^{\ast }.  \label{seed}
\end{eqnarray}%
Here $K_{1}=\Delta _{A}^{-1}K\left( K^{t}K\right) ^{-1}$ and%
\begin{equation}
m(0)=\frac{1}{\left( 2\pi \right) ^{m}}\int_{Z_{B}}\exp \left( iR_{A}Kw-%
\frac{1}{2}w^{t}\alpha w\right) d^{m}w.  \label{inv1}
\end{equation}%
The integrals converge in certain weak sense, i.e. as the integrals of
matrix elements $\int \langle \varphi |\exp \left( iR_{A}Kw\right) \exp
\left( -\frac{1}{2}w^{t}\alpha w\right) |\psi \rangle d^{m}w,$ where $%
\varphi ,\psi $ belong to a dense subspace containing all rapidly decreasing
functions in the Schr\"{o}dinger representation. The relation (\ref{seed})
means that the Gaussian observable $M$ has the structure of a covariant POVM
\cite{aspekty} with the \textquotedblleft core\textquotedblright\ $m(0).$

\section{The basic types}

We will study the possible form of the core\ $m(0)$ for Gaussian
observables. The general case will turn out to be a combination of the three
special cases we first consider separately. The argument proceeds in
parallel to \cite{cj} with $m(0)$ replacing the Choi-Jamiolkowski form of
quantum Gaussian channels.

\textbf{Type1.} Let $Z_{B}=Z_{A},$ so that $m=2s,$ and assume that $K$ hence
$\Delta _{K}$ is nondegenerate. Then $\alpha $ is also nondegenerate by (\ref%
{mudelta}). By making the change of variable $Kw=z$ in (\ref{inv1}) we get
\begin{equation*}
m(0)=\frac{1}{(2\pi )^{2s}\left\vert \det K\right\vert }\int \exp \left(
iR_{A}z\right) \exp \left( -\frac{1}{2}z^{t}\beta \,z\right) d^{2s}z=\frac{%
\left\vert \det K_{1}\right\vert }{\left( 2\pi \right) ^{s}}\rho _{\beta },
\end{equation*}%
where $\beta =\left( K^{-1}\right) ^{t}\alpha K^{-1}$ and $\rho _{\beta }$
is the centered Gaussian density operator with the covariance
matrix $\beta .$ Thus $m(0)$ is a bounded (trace-class) operator. Its
maximal eigenvalue can be found as in \cite{cj} resulting in%
\begin{equation}
\left\Vert \Omega _{\Phi }\right\Vert =\frac{\left\vert \det
K_{1}\right\vert }{\sqrt{\det \left[ \mathrm{abs}\left( \Delta
_{K}^{-1}\alpha \right) +I_{2s}/2\right] }},
\end{equation}%
where $\mathrm{abs}\left( \Delta _{K}^{-1}\alpha \right) $ is the matrix
with eigenvalues equal to modulus of eigenvalues of $\Delta _{K}^{-1}\alpha $
and with the same eigenvectors.

It may be convenient to distinguish the two subtypes of the type 1.

\textbf{Type 1a.} If $\alpha +\frac{i}{2}\Delta _{K}=K^{t}$ $\left( \beta +%
\frac{i}{2}\Delta \right) K$ is nondegenerate, then by theorem 12.23 of \cite%
{QSCI} $\rho _{\beta }$ is a nondegenerate Gaussian density operator. A
special case is the thermal noise state with positive temperature.

\textbf{Type 1b.} If $\alpha +\frac{i}{2}\Delta _{K}$ is maximally
degenerate i.e. $\mbox{rank}\left( \alpha +\frac{i}{2}\Delta _{K}\right) =s$%
, then $\rho _{\beta }$ is pure state, see \cite{cj} (the ground state of
the Hamiltonian $R_{K}\alpha ^{-1}R_{K}^{t}=R_{A}\beta ^{-1}R_{A}^{t}$) and $%
\left\Vert m(0)\right\Vert =\frac{1}{\left( 2\pi \right) ^{s}\left\vert \det
K\right\vert }.$

Type 1a. corresponds to multimode noisy heterodyning with generalized thermal noise, while Type 1b -- to
heterodyning with the minimal quantum (vacuum) noise. Gaussian observables of the type 1 were introduced first
in \cite{h6} (see also the book \cite{aspekty} and references therein). Their classical
capacity was studied in \cite{acc-noJ}  and their entanglement-assisted
capacity was found in \cite{H3}.

\textbf{Type 2.} Let $m\leq s$ with $\alpha >0,$ while $\Delta _{K}=0.$ Then
$R_{K}=R_{A}K$ is the vector operator with commuting selfadjoint components.
The integral (\ref{inv1}) is just the multivariate Gaussian density as a
function of $R_{K}:$
\begin{equation*}
m(0)=\frac{1}{\left( 2\pi \right) ^{s}\sqrt{\det \alpha }}\exp \left( -\frac{%
1}{2}R_{K}\alpha ^{-1}R_{K}^{t}\right) ,
\end{equation*}%
which is a bounded operator. Since the spectrum of $R_{K}$ contains $0,$ we
have $\left\Vert m(0)\right\Vert =\frac{1}{\left( 2\pi \right) ^{s}\sqrt{%
\det \alpha }}.$ In particular, when $m=s$ and $K=\mathrm{diag}\left[1\quad
0 \right]^t _{j=1,\dots ,s},$ we have $R_{K}=\left[ q_{1,\dots ,}q_{s}\right]
$, so we obtain the multimode approximate position measurement (with
correlated Gaussian errors). The noisy homodyning in quantum optics also
belongs to this class. Multimode Gaussian observables of the type 2 were
considered in \cite{hy} where their entanglement-assisted classical capacity
was computed. Notably, the unassisted classical capacity is still an open
problem for this type of observables \cite{hall2}, \cite{hall}.

\textbf{Type 3.} If $\alpha =0\ $then $\Delta _{K}\equiv K^{t}\Delta K=0$
(hence $m\leq s$) by (\ref{mudelta}), and $R_{K}=R_{A}K$ is again the vector
operator with commuting selfadjoint components. Thus we obtain
\begin{equation}
m(0)=\frac{1}{(2\pi )^{m}}\int \exp \left( iR_{K}\,z\right) d^{m}z=\delta
\left( R_{K}\right) ,  \label{delta}
\end{equation}%
where $\delta \left( \cdot \right) $ is Dirac's delta-function. In this case
$m(0)$ is not a bounded operator, but an unbounded nonclosable form.

For example, in the case $m=s$, $K=\mathrm{diag}\left[1\quad 0 \right]%
^t_{j=1,\dots ,s}$ this gives%
\begin{equation*}
\langle \psi |m(0)|\psi ^{\prime }\rangle =\frac{1}{(2\pi )^{s}}\int \langle
\psi |\exp \left( iqx\right) |\psi ^{\prime }\rangle d^{s}x=\langle \psi
|0\rangle \langle 0|\psi ^{\prime }\rangle
\end{equation*}%
for continuous functions $\langle x|\psi \rangle ,\langle x|\psi ^{\prime
}\rangle$ in the Schr\"{o}dinger representation. In particular, multimode
sharp position observable and noiseless homodyning in quantum optics belong
to this type. Gaussian observables of this form were considered in \cite%
{hall} where their classical capacity was found and in \cite{H-QO} where
their entanglement-assisted classical capacity was computed.

Next we will show that in general one can have the combination of the three
types considered above.

\section{Decomposition of a general Gaussian observable}

\label{s2}

Recall that $m=\dim Z_{B}$ and $r_{\alpha }=\mbox{rank}\,\alpha $
is the rank of the $m\times m-$matrix $\alpha .$ The following result is a
generalization of the Williamson's lemma \cite{Williamson39}, cf. \cite{cegh1}.

\begin{lemma}
\label{lem} \emph{Let $\alpha $ be a real symmetric matrix, $\Delta _{K}$ --
a real skew-symmetric matrix such that $\alpha -\frac{i}{2}\Delta _{K}\geq
0. $ Then there is a nondegenerate matrix $T$ such that}
\begin{eqnarray}
\tilde{\alpha} &=&T^{t}\alpha T=\left[
\begin{array}{ccc}
a & 0 & 0 \\
0 & I/2 & 0 \\
0 & 0 & 0%
\end{array}%
\right]
\begin{array}{l}
\}r_{\Delta _{K}} \\
\}r_{\alpha }-r_{\Delta _{K}} \\
\}m-r_{\alpha }%
\end{array}%
,\quad  \label{decom1} \\
\tilde{\Delta}_{K} &=&T^{t}\Delta _{K}T=\left[
\begin{array}{ccc}
\Delta & 0 & 0 \\
0 & 0 & 0 \\
0 & 0 & 0%
\end{array}%
\right] ,  \label{decom2}
\end{eqnarray}%
\emph{where}
\begin{equation*}
\Delta =\mathrm{diag}\left[
\begin{array}{cc}
0 & -1 \\
1 & 0%
\end{array}%
\right] _{j=1,\dots ,r_{\Delta _{K}}/2},\quad a=\mathrm{diag}\left[
\begin{array}{cc}
a_{j} & 0 \\
0 & a_{j}%
\end{array}%
\right] _{j=1,\dots ,r_{\Delta _{K}}/2},
\end{equation*}%
\emph{and} $a_{j}\geq 1/2.$
\end{lemma}

Notice that $r_{\Delta _{K}}=2s_{1}$ is even while $r_{\alpha }$ can be odd.
Denote $s_{2}=r_{\alpha }-r_{\Delta _{K}},s_{3}=m-r_{\alpha }$ the
dimensions of the last two blocks in the decompositions (\ref{decom1}%
), (\ref{decom2}).

Let $\tilde{e}_{j},;\,j=1,\dots ,m$ be the standard basis in $\tilde{Z}_{B}=%
\mathbb{R}^{m}$ in which $\alpha ,\Delta _{K}$ have the block diagonal form (%
\ref{decom1}), (\ref{decom2}) and let $\tilde{Z}_{k}$ be the subspace spanned the vectors $\tilde{e}_{j}$ corresponding to
the $k-$th block in the decompositions, $k=1,2,3.$ Then we have the direct
sum decomposition
\begin{equation}
\tilde{Z}_{B}=\tilde{Z}_{1}\oplus \tilde{Z}_{2}\oplus \tilde{Z}_{3}.
\label{dsum}
\end{equation}

By making the substitution $T^{-1}z=\tilde{z}$ in (\ref{inv1}), we have $%
\tilde{z}=\left[ \tilde{z}_{1},\tilde{z}_{2},\tilde{z}_{3}\right] ^{t}$ and
\begin{equation*}
m(0)=\frac{1}{(2\pi )^{m}\left\vert \det T\right\vert }\int \int \int \exp
\sum_{k=1}^{3}\left( iR_{K}T\tilde{z}_{k}-\frac{1}{2}\tilde{z}_{k}^{t}\alpha
^{(k)}\tilde{z}_{k}\right) d\tilde{z}_{1}d\tilde{z}_{2}d\tilde{z}_{3},
\end{equation*}%
where $\alpha ^{(2)}=I_{s_{2}}/2,$ $\alpha ^{(3)}=0$ and the components of $%
R_{K}T\tilde{z}_{k}$ and $R_{K}T\tilde{z}_{l}$ commute for $k\neq l$ by (\ref%
{decom2}). Hence the exponent under the integral splits into product of
three mutually commuting exponents, and $m(0)$ can be decomposed into the
product of commuting expressions of the types considered in the cases 1-3
above (with possibly odd dimensions for $\tilde{z}_{2},\tilde{z}_{3}$%
):
\begin{equation}
m(0)=\frac{1}{(2\pi )^{m}\left\vert \det T\right\vert }\prod_{k=1}^{3}\int_{%
\tilde{Z}_{k}}\exp \left( iR_{K}T\tilde{z}_{k}-\frac{1}{2}\tilde{z}%
_{k}^{t}\alpha ^{(k)}\tilde{z}_{k}\right) d\tilde{z}_{k}.  \label{prod}
\end{equation}%
In the cases 1,2, where the matrix $\alpha $ is nondegenerate, the
integrals in the product are given by bounded operators, while in the case
3, where the matrix $\alpha $ is zero, the integral is an unbounded form. Hence we
obtain proposition \ref{p1}.


We will need some terminology from the theory of symplectic vector spaces (see e.g. \cite{km}).
Let $L$ be a linear subspace of the symplectic vector space $(Z_{A},\Delta
_{A}).$ Symplectic complement $L^{\perp }$ of $L$ is defined as
\begin{equation*}
L^{\perp }=\left\{ z\in Z_{A}:\Delta _{A}(z,z^{\prime })=0\text{ for all }%
z^{\prime }\in L\right\} .
\end{equation*}%
The subspace is called symplectic if $L\cap L^{\perp }=[0],$ and isotropic
if $L\subseteq L^{\perp }.$ For an isotropic $L$ one has $\dim L\leq \dim Z_{A}/2=s$.
If $L$ is maximal isotropic (Lagrangian), then there is a direct complement of $L$ -- a Lagrangian subspace $L^{\prime}$
such that $Z_{A}=L\oplus L^{\prime}$.

The decomposition (\ref{dsum}) implies%
\begin{equation}
Z_{B}=T(\tilde{Z}_{B})=Z_{1}\oplus Z_{2}\oplus Z_{3},  \label{dsum1}
\end{equation}%
where $Z_{j}=T(\tilde{Z}_{j}),\,j=1,2,3,$ and%
\begin{equation}
Z_{A}\supseteq K(Z_{B})=K(Z_{1})\oplus _{s}K(Z_{2})\oplus _{s}K(Z_{3}),
\label{dsum2}
\end{equation}%
where $\oplus _{s}$ denotes the symplectic direct sum, meaning that the
summands are orthogonal with respect to the form $\Delta _{A}.$ This can
further be complemented to the symplectic direct sum
\begin{equation}
Z_{A}=\hat{Z}_{1}\oplus _{s}\hat{Z}_{2}\oplus _{s}\hat{Z}_{3}\oplus _{s}\hat{%
Z}_{4},  \label{sod}
\end{equation}%
where $\hat{Z}_{1}=K\left( Z_{1}\right) ,\hat{Z}_{2}=K\left( Z_{2}\right)
\oplus \left[ K\left( Z_{2}\right) \right] ^{\prime },\hat{Z}_{3}=K\left(
Z_{3}\right) \oplus \left[ K\left( Z_{3}\right) \right] ^{\prime }$ and $%
\hat{Z}_{4}=\left[ \hat{Z}_{1}\oplus _{s}\hat{Z}_{2}\oplus _{s}\hat{Z}_{3}%
\right] ^{\perp }.$ Here $\hat{Z}_{1}$ is symplectic subspace by
construction, $\dim \hat{Z}_{1}=2s_{1}=r_{\Delta _{K}}$. $K\left(
Z_{2}\right) $ is isotropic subspace which lies in $\left[ \hat{Z}_{1}\right]
^{\perp }$, and $\left[ K\left( Z_{2}\right) \right] ^{\prime }$ is an
isotropic subspace in the direct complement of $K(Z_{B})$ of the same dimensionality $%
s_{2}=r_{\alpha }-r_{\Delta _{K}}$ and such that $\hat{Z}_{2}=K\left(
Z_{2}\right) \oplus \left[ K\left( Z_{2}\right) \right] ^{\prime }$ is
symplectic with $\dim \hat{Z}_{2}=2s_{2}=2\left( r_{\alpha }-r_{\Delta
_{K}}\right)$. $\hat{Z}_{3}$ is built from $K\left( Z_{3}\right) $ in a
similar way and $\dim \hat{Z}_{3}=2s_{3}=2\left( m-r_{\alpha }\right) $. For
this construction to be possible with nonintersecting $\left[ K\left(
Z_{2}\right) \right] ^{\prime }, \left[ K\left( Z_{3}\right) \right]
^{\prime }$ we must have
\begin{equation*}
\dim \left[ K\left( Z_{2}\right) \right] ^{\prime }+\dim \left[ K\left(
Z_{3}\right) \right] ^{\prime }\leq 2s-\dim K(Z_{B}),
\end{equation*}%
or $m-r_{\Delta _{K}}\leq 2s-m,$ the last inequality follows from $%
m-r_{\Delta _{K}}\leq s-r_{\Delta _{K}}/2$ because the dimensionality of any
isotropic subspace in $Z_{A}$ is $\leq s.$ Thus $\hat{Z}_{1}\oplus _{s}\hat{Z%
}_{2}\oplus _{s}\hat{Z}_{3}$ is symplectic subspace of dimensionality
\begin{equation*}
2s_{1}+2s_{2}+2s_{3}=r_{\Delta _{K}}+2\left( r_{\alpha }-r_{\Delta
_{K}}\right) +2\left( m-r_{\alpha }\right) =2m-r_{\Delta _{K}}\leq 2s,
\end{equation*}%
Hence it has the symplectic complement $\hat{Z}_{4}$ which is either $\left[
0\right] $ or symplectic. By construction, the subspaces $\hat{Z}%
_{j},\,j=1,2,3,4$ are mutually symplectic orthogonal, so the product (\ref%
{prod}) can be further transformed into tensor product in the space $%
\mathcal{H}_{A}$ .

\begin{lemma}
Denote $s_{C}=r_{\alpha }-r_{\Delta _{K}}/2.$ Let $\alpha $ satisfy (\ref%
{mudelta}), then there exists $(s_{C}\times s_{C})-$matrix $\alpha _{C}\geq
\pm \frac{i}{2}\Delta _{C}$ satisfying (\ref{proj}), namely%
\begin{equation*}
K^{t}P^{t}\Lambda ^{t}\alpha _{C}\Lambda PK=\alpha ,
\end{equation*}%
where $\Lambda $ is $2s_{C}\times 2s_{C}-$matrix defined in (\ref{lam})
below.
\end{lemma}

\emph{Proof}. Define the basis in $\hat{Z}_{1}=KT(\tilde{Z}_{1})\subseteq
Z_{A}$ as follows:%
\begin{equation*}
\hat{e}_{j}=K\dot{T}\,\tilde{e}_{2j-1},\quad \tilde{h}_{j}=K\dot{T}\,\tilde{e%
}_{2j};\,\quad j=1,\dots ,r_{\Delta _{K}}/2.
\end{equation*}%
Then according to (\ref{decom2}) it is symplectic
\begin{eqnarray*}
\Delta (\hat{e}_{j},\hat{e}_{k}) &=&\tilde{\Delta}_{K}(\tilde{e}_{j},\tilde{e%
}_{k})=0;\quad  \\
\Delta (\hat{e}_{j},\tilde{h}_{k}) &=&\tilde{\Delta}_{K}(\tilde{e}_{j},%
\tilde{h}_{k})=\delta _{jk};\quad \,j,k=1,\dots ,r_{\Delta _{K}}/2.
\end{eqnarray*}

Further, consider the basis $\hat{e}_{j}=K\dot{T}\,\tilde{e}_{j+r_{\Delta
_{K}}/2};\,\,j=r_{\Delta _{K}}/2+1,\dots ,r_{\alpha }-r_{\Delta _{K}}/2=s_{C}
$  in $K\left(
Z_{2}\right) $ and complement it by the basis $\tilde{h}_{k};k=r_{\Delta
_{K}}/2+1,\dots ,r_{\alpha }-r_{\Delta _{K}}/2$ in $\left[ K\left(
Z_{2}\right) \right] ^{\prime }$ such that $\left\{ \hat{e}_{j},\tilde{h}%
_{k}\right\} $ is symplectic basis in $\hat{Z}_{2}=K\left( Z_{2}\right)
\oplus \left[ K\left( Z_{2}\right) \right] ^{\prime }.$ Thus
\begin{equation*}
\left\{ \hat{e}_{j},\tilde{h}_{k};\,j,k=1,\dots ,r_{\alpha }-r_{\Delta
_{K}}/2=s_{C}\right\}
\end{equation*}%
becomes a symplectic basis in the subspace
\begin{equation}
Z_{C}\equiv \hat{Z}_{1}\oplus _{s}\hat{Z}_{2}\subseteq Z_{A}  \label{zc}
\end{equation}%
supplied with the symplectic form $\Delta _{C}$ which is restriction of $%
\Delta _{A}$ to $Z_{C}.$

Defining the involution $\Lambda $ in $Z_{C}$ by%
\begin{equation}
\Lambda \hat{e}_{j}=\hat{e}_{j},\quad \Lambda \hat{h}_{j}=-\hat{h}_{j};\quad
j=1,\dots ,r_{\alpha }-r_{\Delta _{K}}/2=s_{C}.  \label{lam}
\end{equation}%
and the projection $P$ from $Z_{A}$ to $K(Z_{1})\oplus _{s}K(Z_{2})$:
\begin{eqnarray}
P\hat{e}_{j} &=&\hat{e}_{j};\quad j=1,\dots ,s_{C},\quad P\hat{h}_{j}=\hat{h}%
_{j};\quad j=1,\dots ,r_{\Delta _{K}}/2 \notag \\
Pz_{A} &=&0;\quad z_{A}\in \left[ K\left( Z_{2}\right) \right] ^{\prime
}\oplus _{s}\hat{Z}_{3}\oplus _{s}\hat{Z}_{4},\label{pr}
\end{eqnarray}%
we have $\Lambda \Delta _{C}\Lambda =-\Delta _{C}$ and
\begin{equation}
K^{t}P^{t}\Lambda \Delta _{C}\Lambda PK=-\Delta _{K}.  \label{com}
\end{equation}%
Thus the commutator matrix $\Delta _{K}^{C}$ of the observables $%
R_{C}\Lambda PK$ is equal to $-\Delta _{K}$ implying that the commutators of
the components of vector observable $X_{B}$ in (\ref{selfa}) are zeroes. Hence
they have the joint spectral measure $E_{AC}(d^{m}z)$.

Define the $s_{C}\times s_{C}-$matrix $\alpha _{C}$ by the matrix elements
\begin{eqnarray*}
\alpha _{C}(\Lambda \hat{e}_{j},\Lambda \hat{e}_{k}) &=&\alpha _{C}(\Lambda
\tilde{h}_{j},\Lambda \tilde{h}_{k})=a_{j}\delta _{jk}, \\
\alpha _{C}(\Lambda \hat{e}_{j},\Lambda \tilde{h}_{k}) &=&0;\quad
\,j,k=1,\dots ,r_{\alpha }-r_{\Delta _{K}}/2=s_{C},
\end{eqnarray*}%
where we put $a_{j}=1/2$ for $j=r_{\Delta _{K}}/2+1,\dots ,r_{\alpha
}-r_{\Delta _{K}}/2=s_{C}.$ Then it satisfies $\alpha _{C}\geq \pm \frac{i}{2}%
\Delta _{C}$ implying that there is centered Gaussian state $\rho _{C}$
with the covariance matrix $\alpha _{C}.$ Further, $T^{t}K^{t}P^{t}\Lambda
\alpha _{C}\Lambda KPT=\tilde{\alpha},$ so that $K^{t}P^{t}\Lambda \alpha
_{C}\Lambda PK=\alpha $ which means (\ref{proj}) .

This accomplishes the construction of the quantum ancilla $C$ and the spectral measure
$E_{AC}$, and hence the proof of theorem \ref{t1}.  $\square $

\textbf{Remark}. From the construction above one can see also that if a hybrid (quantum-classical) ancilla is allowed
then it can have $r_{\Delta _{K}}/2=s_{1}$ quantum modes (based on the subspace $K\left( Z_{1}\right)$)
and $r_{\alpha}-r_{\Delta _{K}}/2=s_{2}$ classical dimensions (of the subspace $K\left( Z_{2}\right)$).

\end{document}